\input{aipcheck.tex}

\documentclass{aipproc}

\newcommand{\ri}{\ensuremath{n}}
\newcommand{\unts}[1]{\ensuremath{\,\mathrm{#1}}}
\newcommand{\HI}{H\,\textsc{i}}

\newcommand{\dem}{DEM L\,316}

\newcommand{\apj}{Ap. J.,}
\newcommand{\apjl}{Ap. J. Lett.,}

\newcommand{\nat}{Nature,}

\newcommand\doingARLO[2][]{%
  \ifx\mmref\undefined #1\else #2\fi
} \layoutstyle{8x11double}

\SetInternalRegister\hbadness{8000} 

\begin{document}

\title
      [GRB Remnants ]
      {GRB Remnants }


\author{Tsvi Piran and Shai Ayal}{
  address={Racah Institute for Physics, The Hebrew University, Jerusalem, Israel 91904},
  email={tsvi@phys.huji.ac.il}
  thanks={This work was supported by the US-Israel BSF}
}

\begin{abstract}

The realization that GRBs are narrowly beamed implied that the
actual rate of GRBs is much larger than the observed one. There
are 500 unobserved GRBs for each observed one.  The lack of a
clear trigger makes it hard to detect these unobserved GRBs as
orphan afterglows. At late time, hundreds or thousands of years
after a GRB, we expect to observe a GRB remnant (GRBR). These
remnants could be distinguished from the more frequent SNRs using
their different morphology. While SNRs are spherical, GRBRs that
arise from a highly collimated flow, are expected to be initially
nonspehrical. We ask the question for  how long can we identify a
GRBR among the more common SNRs?  Using SPH simulations we follow
the evolution of a GRBR and calculate the image of the remnant
produced by bremsstrahlung and by synchrotron emission. We find
that the GRBR  becomes spherical after $ \sim  3000 {\rm
yr}(E_{51}/n)^{1/3}$ at $R\sim 12{\rm pc} (E_{51}/n)^{1/3}$,
where $E_{51}$ is the initial energy in units of $10^{51}{\rm
erg}$ and $n$ is the surrounding ISM number density in ${\rm
cm}^{-3}$. We expect $0.5 (E_{51}/n)^{1/3}$ non-spherical GRBs
per galaxy. Namely, we expect $\sim ~ 20$ non spherical GRBRs
with angular sizes $\sim \mu$arcsec within a distance of 10Mpc.
These results are modified if there is an underlying spherical
supernova. In this case the GRBR will remain spherical only for
$\sim150 {\rm yr} (E_{51}/n)^{1/3}$ and the number of
non-spherical GRBRs is smaller by a factor of 10 and their size
is smaller by a factor of 3.

\end{abstract}

\date{\today}

\maketitle

\section{Introduction}

A $\gamma$-ray burst (GRB) that originates within a galactic disk
deposits $\sim 10^{51}$ergs into the ISM. This results in  a
blast wave whose initial phase produces the afterglow. The late
phase of the blast wave evolution would result, as noted by
Chevalier \cite{chevalier74} in the context of supernova remnants
(SNRs), in a cool expanding \HI~shell. The shell will remain
distinct from its surrounding until it has slowed down to a
velocity of $\approx 10\unts{km\,s^{-1}}$ \citep{loeb98}, which
should happen within $2.3\cdot10^6\unts{yr}~E_{51}^{0.32}$ where
$E_{51}$ is the initial energy in units of $10^{51}\unts{erg}$.

The observed rate of GRBs is one per $\sim10^7\unts{yr}$ per
galaxy \citep{schmidt99}. The implied GRB isotropic energy is of
the order of $10^{53}$ergs.  These estimates suggested that there
are a few remnants per galaxy at any given time. As it was
believed that the GRB explosions were much more energetic than
SNs, Loeb and Perna \citep{loeb98} suggested that GRBRs would
form HI supershells. This giant structures require much more
energy than what a usual SN can supply.

However, the realization that GRBs are beamed
\citep{sari99,halpern99,sari99a,kulkarni99,harrison99} changed
both estimate. First the rate of GRBs is much higher. Beamed
GRBs  illuminate only a fraction $f_b$ of the sky, thus their
rate should be higher by a factor of $f_b^{-1}$. With $f_b\sim
0.002$ \citep{Frail01} we  expect several thousand GRB remnants
per galaxy. On the other hand the energy output of each GRB is
much smaller \cite{Frail01,Piran01,PK01}. Thus they cannot produce
the giant HI shells.

How can we distinguish a GRBR from and SNR and for how long? Both
GRBs and SNs deposit a comparable kinetic energy ($\sim
10^{51}\unts{erg}$) into the ISM. The energy injection in a GRB
is in a form two narrow relativistic beams containing $\sim
10^{-5}M_\odot$. A SN deposits this energy spherically with $\sim
10 M_\odot$. In both cases the  expected late evolution is
similar. At this late stage
 both  remnants are in the Sedov \citep{Sedov59}
regime where all the kinetic energy is in the ejecta and all the
mass is in the surrounding ISM. A key distinguishing feature
unique to GRB remnants could be their beamed nature. We expect
that the beamed emission would  lead to a distinct double shell
morphology at intermediate times. The late time behavior of the
GRB remnant is expected to be spherical. To establish how many
\HI~shells are GRB remnants we need to find out the expected
morphology of GRB remnants and how long they stay non-spherical
and distinguishable from SNRs. Establishing how many of the
\HI~shells are GRB remnants would make it possible to directly
estimate the local rate of GRBs, determine $\epsilon$, the
efficiency of converting the explosion energy into $\gamma$-rays,
and the beaming factor $f_b$ \citep{loeb98}.

We model the intermediate evolution of a beamed GRB by two blobs
of dense material moving into the ISM in opposite directions. We
follow numerically the hydrodynamic evolution \citep{Ayal_Piran}.
We find that the morphology depends on the dimensionless ratio
between the accumulated mass and the initial kinetic energy, $\mu
\equiv Mc^2/E_0$. When  $\mu \sim 2.1\times10^5$ (at $t \sim 3000
{\rm yr}(E_{51}/n)^{1/3}$ and $R\sim 12{\rm pc}
(E_{51}/n)^{1/3}$) ,the remnant becomes spherical and
indistinguishable from a SNR.

An additional complication arises if the GRB is accompanied by a
supernova, as suggested in the Collapsar model
\cite{Woosley93,Pac98,MacFadyen_W99}. The supernova produces an
underlying massive spherical Newtonian shell that propagates
outwards. At $\mu \sim 3000$ corresponding to   $t \sim150 {\rm
yr} (E_{51}/n)^{1/3}$ and $R~4{\rm pc} (E_{51}/n)^{1/3}$ this
shell will catch  the non-spherical GRBR and the system will
quickly become spherical. In this case the number of
non-spherical GRBRs is smaller by a factor of 10 and their size is
smaller by a factor of 3.

\section{The Numerical Simulations}
\subsection{The Model}
\begin{figure}
    \resizebox{.9\columnwidth}{!}
  {\includegraphics{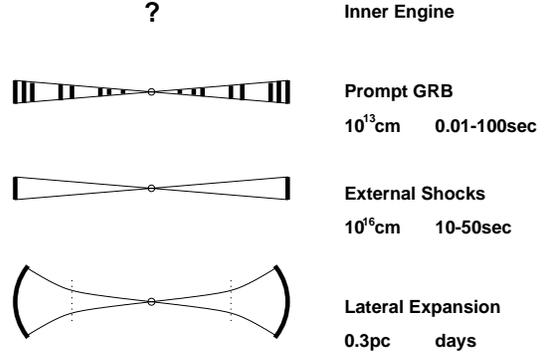}}
    \caption{
     A schematic evolution of a GRB during its relativitistic phase.
     From top to bottom:
     (a) An inner engine accelerates relativistic jets. (b) Collisions within the jets
     produce the observed GRB. (c) External shocks produce the afterglow. (d)
     The jets expand sideways as they slow down.}
    \label{fig:scheamtic1}
\end{figure}

A GRB occurs when    a compact `inner engine' ejects two
ultra-relativistic beams . Internal collisions within these beams
leads to the GRB (See a schematic description in Fig.
\ref{fig:scheamtic1}.). Later external shocks caused by collisions
with circumstellar matter produce the afterglow. The matter slows
down during this interaction and its bulk Lorenz factor $\Gamma$,
decreases. The ejecta stays collimated only until $\Gamma$ drops
below $\sim 1/\theta_0$, at approximately $2.9
\unts{hr}(E_{51}/n_{\rm I})^{1/3}(\theta_0/0.1)^{8/3}$ after the
GRB \citep{rhoads97,sari99} where $\theta_0$ is the initial
angular width.
 At this time the matter starts expanding
sideways causing, for an adiabatic evolution, an exponential
slowing down \citep{rhoads97}. The ejecta continues to expand
sideways at an almost constant radial distance from the source
$R_0\sim 0.3\unts{pc}~E_{51}^{1/3}\ri^{-1/3}$  until it becomes
non-relativistic. At this stage, we begin our simulation.

\begin{figure}
    \resizebox{.9\columnwidth}{!}
  {\includegraphics{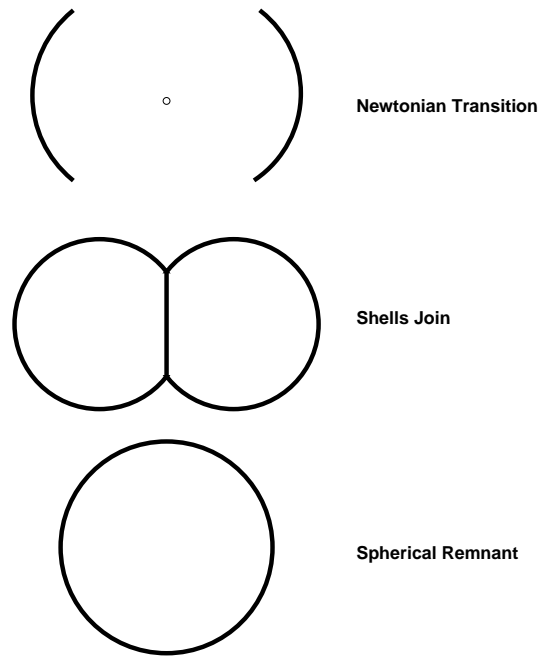}}
    \caption{
     A schematic evolution of a GRBR. From top to bottom:
     (a) Initial conditions around the Newtonian transition. (b) Shells
     collision along the equatorial plane. (c) A late time spherical
     shell.}
    \label{fig:scheamtic2}
\end{figure}

Without a detailed numerical modeling of the relativistic phase of
the ejecta we have only an approximate description of the initial
conditions. We expect the angular width of the ejecta to be $\sim
1\unts{rad}$ and we are constrained by the energy conservation:
\begin{equation}
R_0 \sim 0.3 \unts{pc} E_{51}^{1/3}\ri^{-1/3}(v_0/c)^{-2/3}
\,.\label{eq:const}
\end{equation}
Our initial conditions comprise two identical blobs moving at
$v_0\sim c/3$ in opposing directions into the ISM.  Both the
blobs and the ISM are modeled by a cold $\gamma=5/3$ ideal gas.
The blobs are are denser than the ISM.

Luckily  the intermediate and late evolution of the ejecta are
insensitive to the  initial conditions. Already in the
intermediate stage  we are in the Sedov regime, the mass is
dominated by the ``external'' ISM gas which washes out any
variations in the initial conditions of the ejecta. Our numerical
simulations \cite{Ayal_Piran} verified this expectation and
different initial densities, angular widths and shapes of the
blobs led to essentialy similar late time configurations.

Our code  is based on the Newtonian version of the smooth
particle hydrodynamics (SPH) code introduced in \cite{sphpn}. The
code was adapted for the specific problem at hand. We have also
used the post Newtonian  version of the code to take account of
possible initial relativistic effects  (with an initial blob
velocity of $c/3$).

\begin{figure}[htbp]
  \resizebox{.9\columnwidth}{!}
  {\includegraphics{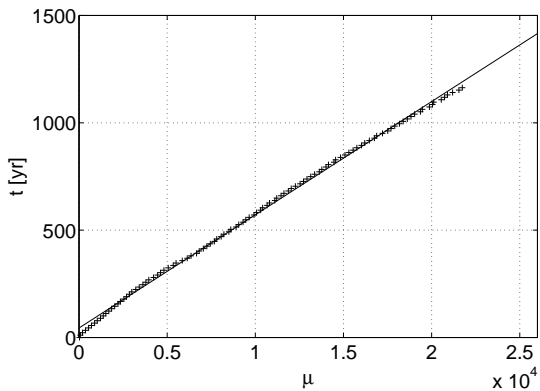}}
  \caption{Time as a function of $\mu$. The linear relation between time
      and $\mu$ is $t\sim 0.046\mu(E_{51}/\ri)^{-1/3} \unts{yr}$.}\label{fig:mu}
\end{figure}

Once we choose the initial velocity. Equation (\ref{eq:const})
leaves us with the freedom of choosing two out of the three
parameters $E_0$, $R_0$ and $\ri$, the initial energy, distance
and ISM density respectively. In presenting the results we choose
$E_0$ and $n$. To parameterize the evolution of the remnant we
utilize the fact that mass scales linearly with initial energy
and define the dimensionless parameter $\mu=Mc^2/E_0$ where $M$
is the accumulated shell mass. We define $M$ as all mass with
density above $2n$. We show all subsequent results as functions of
$\mu$. The simulations begin with $\mu\sim 1$. Conveniently $\mu$
scales linearly with time with: $t\sim
0.046\mu(E_{51}/\ri)^{-1/3} \unts{yr}$, as can be seen from Fig.
\ref{fig:mu}.

\subsection{Results} \label{sec:Res}

As each blob collides with the ISM it produces a bow shock. This
shock propagates also in the direction perpendicular to the blob's
velocity. As the shocked blob material heats up it begins to
expand backwards and a backwards going shock develops as well.
The expected morphology of the remnant will therefore be of two
expanding shells which will eventually join, producing yet
another shock. At late times the shells  merge and become a single
spherical shell.

\begin{figure}
    \resizebox{.9\columnwidth}{!}
  {\includegraphics{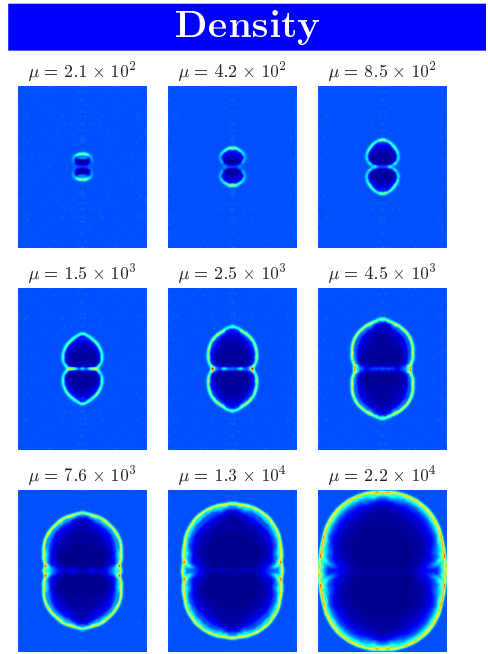}}
    \caption{
      Equally spaced density contours
      ($\rho=1.5n,\,2n,\,\ldots,\,3.5n$) at
      $\mu=9.5\times10^2,\,1.4\times10^3,\, 2.4\times10^3,\,
      3.9\times10^3,\, 6.1\times10^3,\, 10^4,\, 1.8\times10^4,\,
      3\times10^4,\, 5\times10^4$ (left to right, top to bottom)}
    \label{fig:dens_cont}
\end{figure}

 Fig. \ref{fig:scheamtic2}
depicts the expected schematic hydrodynamic behaviour after the
Newtonain transition. This is indeed confirmed in the
computation. Fig.~\ref{fig:dens_cont}, depicts the density
contours along the evolution. We observe the expected evolution:
from two individual blobs via a peanut shape configuration with a
shock along the equator towards a more and more spherical
configuration at late times.

\begin{figure}
 \resizebox{.9\columnwidth}{!}
  {\includegraphics{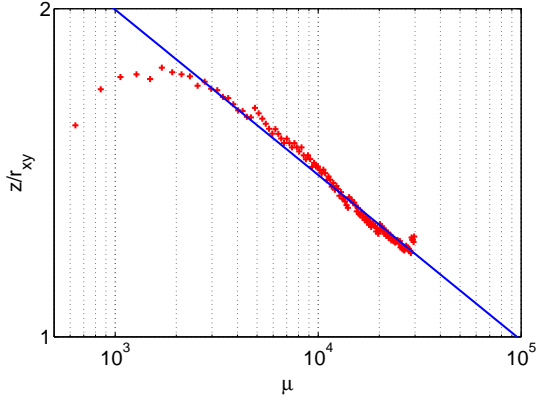}}
    \caption{The ratio between the radius $r_{xy}$ of the shock and the $z$
      position of the shock. The solid line is the best fit power law
      $\mu^{-0.15}$. The ratio will reach a value of 1 at $\mu\sim
      2.1\times 10^5\unts{yr}$}
    \label{fig:ratio}
\end{figure}

The ratio $z_{\rm max}/r_{xy}$ can  be approximated by a power
law as shown in Fig~\ref{fig:ratio}. In our simulation this ratio
is always between 1 and 2 so that the power law fit is very
inaccurate. This ratio decreases in time as a power law with an
exponent of $-0.15$. Extrapolating this power law we see that
this ratio reaches a value of 1 at $\mu\sim 2.1\times 10^5$. At
this time the shock has a spherical shape with $z=r_{xy}\sim
15(E_{51}/\ri)^{1/3} \unts{pc}$. Even then the shock will not be
completely spherically symmetric as there would still be a ring of
shocks around the ``equator'' where the shells have collided.

\begin{figure}
 \resizebox{.9\columnwidth}{!}
  {\includegraphics{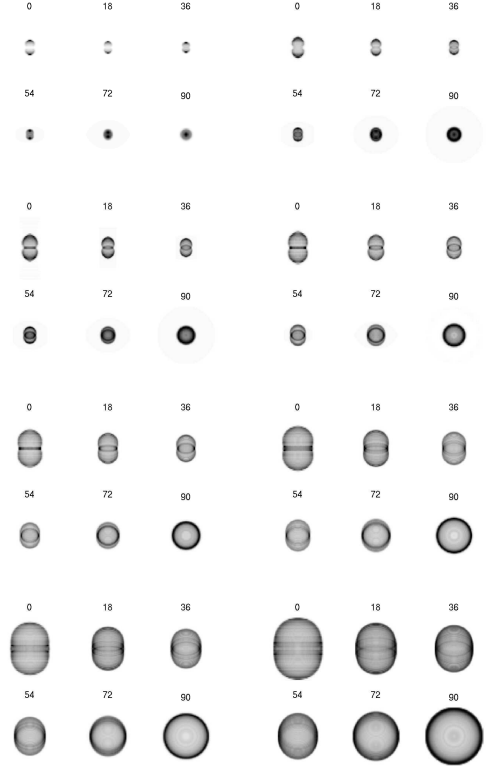}}
    \caption{Images of the remnant, bremsstrahlung emission. The
      number above each image is the angle of inclination in degrees.
      The images are shown at the same $\mu$ as the last 8 panels of
      figure~\ref{fig:dens_cont}.  }
    \label{fig:images_brem}
\end{figure}

\begin{figure}
 \resizebox{.9\columnwidth}{!}
  {\includegraphics{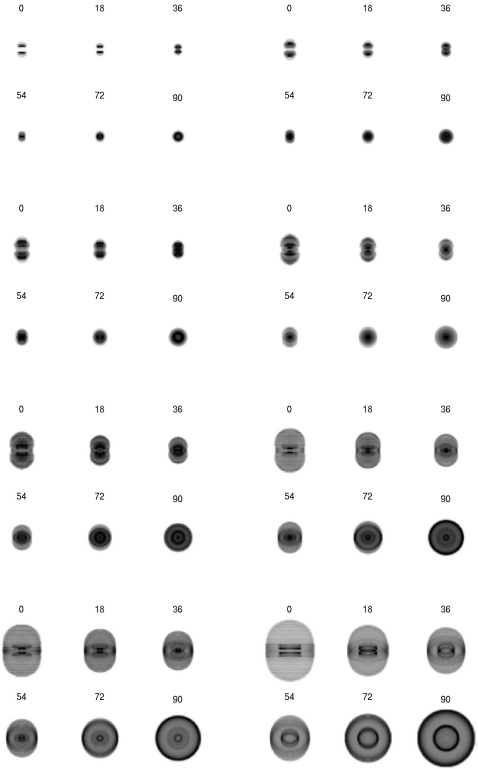}}
    \caption{Images of the remnant, synchrotron emission. The $\mu$ are
      the same as the last 8 panels in figure \ref{fig:images_brem}}
    \label{fig:images_sync}
\end{figure}

\begin{figure}
 \resizebox{.9\columnwidth}{!}
  {\includegraphics{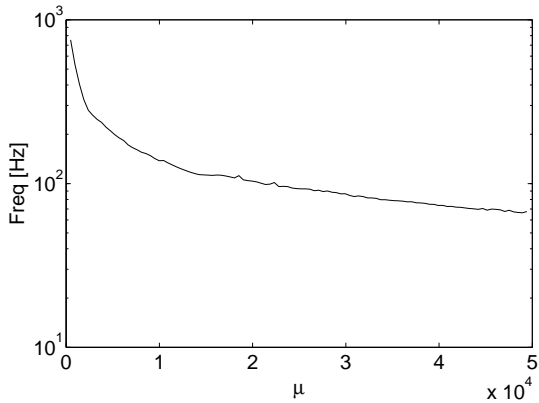}}
    \caption{
      The characteristic synchrotron frequency as a function of $\mu$.}
    \label{fig:freq_s}
\end{figure}

\begin{figure}
 \resizebox{.9\columnwidth}{!}
  {\includegraphics{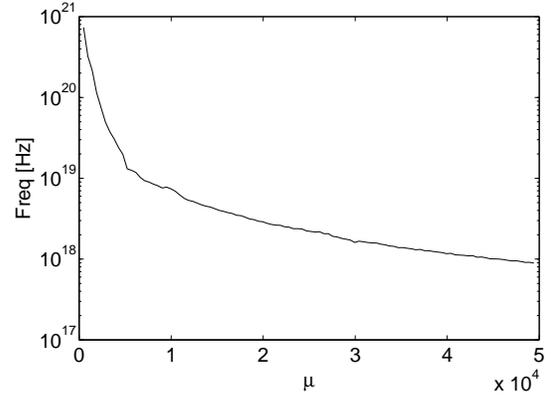}}
    \caption{The characteristic bremsstrahlung frequency  as a function of
      $\mu$.}
    \label{fig:freq_b}
\end{figure}

Figures \ref{fig:images_brem} and \ref{fig:images_sync} depict the
images of the remnant as a function of time and angles of
inclination. We show images due to bremsstrahlung emission and
synchrotron emission. The images are constructed assuming that
all the gas is optically thin in the relevant frequencies. The
bremsstrahlung luminosity (Fig. \ref{fig:images_brem}) was
calculated assuming that the volume emissivity is proportional to
$\rho^2\varepsilon^{1/2}$ \citep{lang80}. In calculating the
synchrotron emissivity (Fig. \ref{fig:images_sync}) we assumed
that both the magnetic field energy density and the number
density of the relativistic electrons are proportional to the
internal energy density of the gas with constant proportionality
factors $\epsilon_B$ and $\epsilon_e$ respectively. We further
assume that the relativistic electron number density is a power
law in energy. Under these assumptions the volume emissivity is
proportional to $\rho^2\varepsilon^2$ \citep[e.g.][]{shu91}. In
the late images there are two bright circles at the lines where
the colliding blobs form a hot shocked region. In figures
\ref{fig:freq_s} and \ref{fig:freq_b} we show the characteristic
emission frequencies. For bremsstrahlung this is $kT/h$ where $T$
is the temperature of the gas. For synchrotron emission we assume
$\epsilon_B=0.1$. The characteristic frequency in this case is
the Larmor  frequency $eB/m_e$ where $e$, $B$ and $m_e$ are the
electron charge, the magnetic field and the electron mass
respectively.

\section{Discussion}

The long time shape of a GRB remnant is insensitive to the exact
initial morphology, angular width and density of the ejecta.
Initially the remnant is highly non-spherical. It becomes
spherical as time advances and the ratio between its height and
radius  approaches unity when
\begin{equation}
\mu \approx 2.1\times 10^5 \ .
\end{equation}
This corresponds to
\begin{equation}
 t \sim 10^4  \unts{yr} ~E_{51}^{1/3}\ri^{-1/3} \ ,
\end{equation}
and
\begin{equation}
R\sim 12{\rm pc} (E_{51}/n)^{1/3}  \ .
\end{equation}
 After this time it will be difficult to distinguish a GRB
remnant from a SNR on the basis of its morphology alone.

Using as the observed GRB rate $R_{GRB} =
10^{-7}\unts{yr^{-1}\,gal^{-1}}$ \cite{schmidt99} the expected
number of non-spherical GRBRs  per galaxy is:
\begin{equation}
0.5 \unts{yr^{-1}} ({f_b \over 500})^{-1} ({R_{GRB} \over
10^{-7}})
 E_{51}^{1/3}\ri^{-1/3} \ .
\end{equation}
This value depends of course critically on the typical beaming
factor, $f_b$.  It should be compared with the expectation of 100
similar aged ($10^4$ yrs) SNRs per galaxy. We would expect 20 non
spherical GRBRs up to a distance of 10 Mpc. The angular sizes of
these GRBRs would be around a $\mu$arcsec.

\subsection{Implications to \dem}

\begin{figure}
 \resizebox{.9\columnwidth}{!}
  {\includegraphics{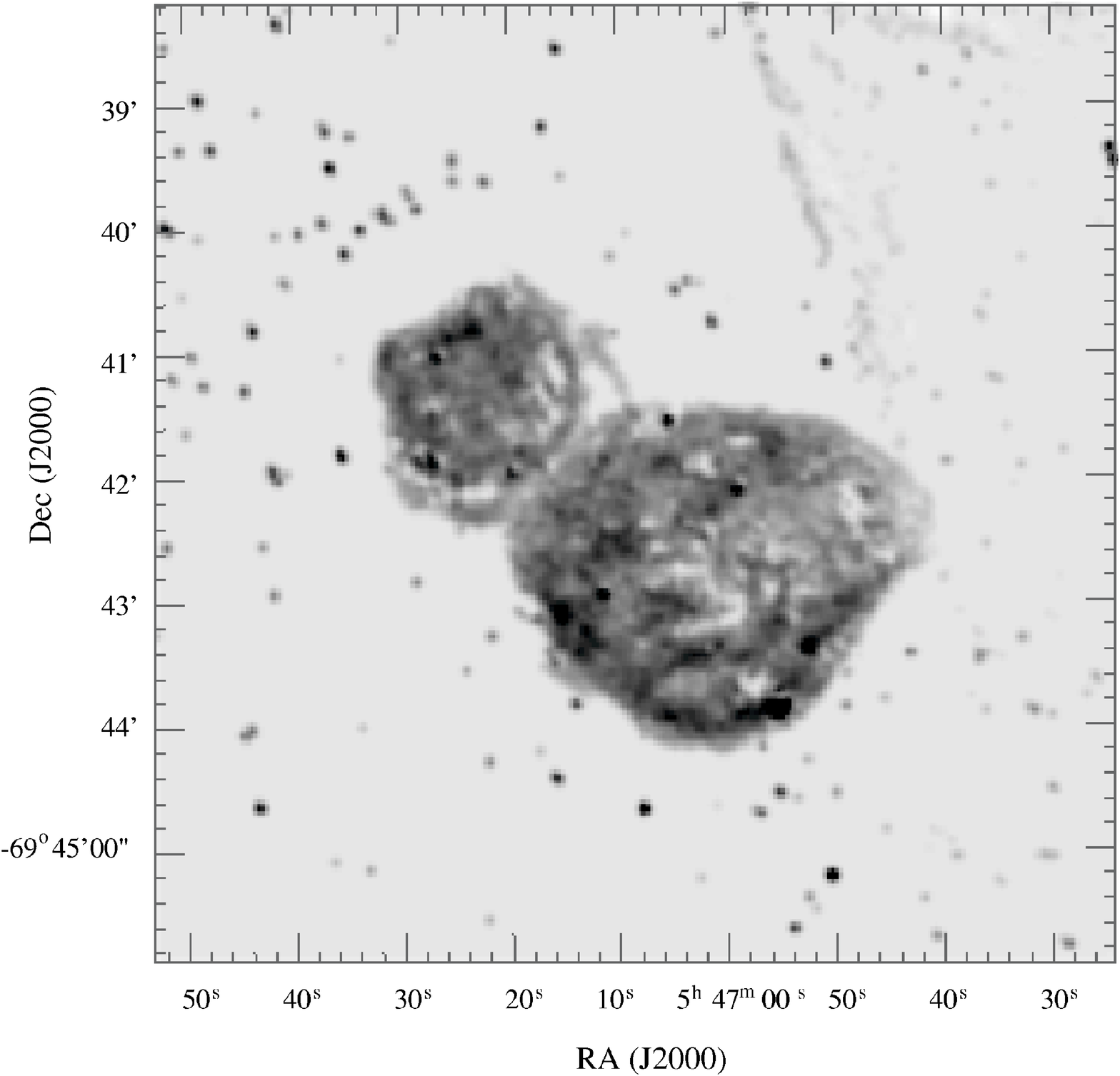}}
    \caption{
      \dem~  in the LMC (from \citep{Williamsetal97}).}
    \label{fig:dem}
\end{figure}

\dem~\citep{Williamsetal97} in the LMC looks like two colliding
bubbles (see Fig. \ref{fig:dem}) . It is thought to result from a
collision between  two SNRs. This requires, of course, an unlikely
coincidence in the timing and the location of the two SNes. An
interesting possibility is that \dem~ is a GRBR. Does this fit our
model? \dem~is far from spherical and has a distinct double shell
morphology, most similar to our results at $\mu\sim 10^4$ (see
Fig.~\ref{fig:dens_cont}). The  $\mu$ ratio measured for \dem~is
$\approx 7\times10^5$. However, according to our results this is
far after the spherical transition. A GRB remnant would already
be spherical at this stage.  This discrepancy rules out the
possibility of fitting \dem~ with our model for a GRB remnant.

\subsection{An Underlying Supernova}

Our model should be modified if the GRB relativistic beams are
accompanied by an underlying spherical supernova, as would be
expected in the Collapsar model
\cite{Woosley93,Pac98,MacFadyen_W99}. In this case a spherical
shell of $\sim 10 m_\odot$, the supernova ejecta, will accompany
the GRB beams. This ejecta propagates at a much lower, Newtonain
velocity, with initial values of $\sim 10^{4}$km/sec. However, it
will not slow down while the outer GRB ejecta is piling up the
external matter and is slowed down. Eventually it will catch the
GRB ejecta.  It is clear that at this stage the SN shell would
tend to make the GRBR bow shock more spherical. To determine when
this will happen one needs another set of numerical simulations.
These are in progress now. However, we can attempt to estimate
when the slower SN ejecta will catch up the slowing down GRBR
remnant. Assuming that the SN ejecta does not slow down (as the
GRBR ejecta clears the surrounding ISM matter) we find that this
will happen at: $\mu \approx  3000$, namely at   $t \approx
150{\rm yr} (E_{51}/n)^{1/3}$and $R \approx 4 {\rm pc}
(E_{51}/n)^{1/3}$. This happens around the time that the two
shells collide on the equator. We expect to see an enhanced
emission due to this collision and then the system will become
quickly spherical. The expected number of non spherical GRBRs and
their corresponding sizes  would be smaller by a factor of 10
then the values estimated earlier for the simple evolution of the
beamed GRB ejecta. Thus we expect one or two non spherical GRBRs
with distances up to 20 Mpc and their sizes would be around 0.1
 $\mu$arcsec.

\begin{theacknowledgments}

This research was supported by a grant from the US-ISRAEL BSF.

\end{theacknowledgments}


\doingARLO[\bibliographystyle{aipproc}]
          {\ifthenelse{\equal{\AIPcitestyleselect}{num}}
             {\bibliographystyle{arlonum}}
             {\bibliographystyle{arlobib}}
          }
\bibliography{sample}

\begin{thebibliography}{25}



\bibitem[{{Chevalier}(1974)}]{chevalier74} {Chevalier}, R.~A. 1974, \apj~ 188, 501

\bibitem[{{Loeb} \& {Perna}(1998)}]{loeb98}
{Loeb}, A. \& {Perna}, R. 1998, \apjl~ 503, L35

\bibitem[{{Schmidt}(1999)}]{schmidt99}
{Schmidt}, M. 1999, \apjl~ 523, L117

\bibitem[{{Sari} {et~al.}(1999){Sari}, {Piran}, \& {Halpern}}]{sari99}
{Sari}, R., {Piran}, T., \& {Halpern}, J.~P. 1999, \apjl~ 519, L17


\bibitem[{{Halpern} {et~al.}(1999){Halpern}, {Kemp}, {Piran}, \&
  {Bershady}}]{halpern99}
{Halpern}, J.~P., {Kemp}, J., {Piran}, T., \& {Bershady}, M.~A.
1999, \apjl~
  517, L105

\bibitem[{{Sari}(1999)}]{sari99a}
{Sari}, R. 1999, in Proc. of the 5th Huntsville Gamma-Ray Burst
Symposium

\bibitem[{{Kulkarni} {et~al.}(1999){Kulkarni}, {Djorgovski}, {Odewahn},
  {Bloom}, {Gal}, {Koresko}, {Harrison}, {Lubin}, {Armus}, {Sari},
  {Illingworth}, {Kelson}, {Magee}, {van Dokkum}, {Frail}, {Mulchaey},
  {Malkan}, {McClean}, {Teplitz}, {Koerner}, {Kirkpatrick}, {Kobayashi},
  {Yadigaroglu}, {Halpern}, {Piran}, {Goodrich}, {Chaffee}, {Feroci}, \&
  {Costa}}]{kulkarni99}
{Kulkarni}, S.~R., {Djorgovski}, S.~G., {Odewahn}, S.~C.,
{Bloom}, J.~S.,
  {Gal}, R.~R., {Koresko}, C.~D., {Harrison}, F.~A., {Lubin}, L.~M., {Armus},
  L., {Sari}, R., {Illingworth}, G.~D., {Kelson}, D.~D., {Magee}, D.~K., {van
  Dokkum}, P.~G., {Frail}, D.~A., {Mulchaey}, J.~S., {Malkan}, M.~A.,
  {McClean}, I.~S., {Teplitz}, H.~I., {Koerner}, D., {Kirkpatrick}, D.,
  {Kobayashi}, N., {Yadigaroglu}, I.~., {Halpern}, J., {Piran}, T., {Goodrich},
  R.~W., {Chaffee}, F.~H., {Feroci}, M., \& {Costa}, E. 1999, \nat~ 398, 389


\bibitem[{{Harrison} {et~al.}(1999){Harrison}, {Bloom}, {Frail}, {Sari},
  {Kulkarni}, {Djorgovski}, {Axelrod}, {Mould}, {Schmidt}, {Wieringa}, {Wark},
  {Subrahmanyan}, {McConnell}, {McCarthy}, {Schaefer}, {McMahon}, {Markze},
  {Firth}, {Soffitta}, \& {Amati}}]{harrison99}
{Harrison}, F.~A., {Bloom}, J.~S., {Frail}, D.~A., {Sari}, R.,
{Kulkarni},
  S.~R., {Djorgovski}, S.~G., {Axelrod}, T., {Mould}, J., {Schmidt}, B.~P.,
  {Wieringa}, M.~H., {Wark}, R.~M., {Subrahmanyan}, R., {McConnell}, D.,
  {McCarthy}, P.~J., {Schaefer}, B.~E., {McMahon}, R.~G., {Markze}, R.~O.,
  {Firth}, E., {Soffitta}, P., \& {Amati}, L. 1999, \apjl~ 523, L121

\bibitem[{Frail} {et~al.}(2001)]{Frail01}{{Frail}, D.~A. and {Kulkarni}, S.~R. and {Sari}, R. and {Djorgovski}, S.~G. and
    {Bloom}, J.~S. and {Galama}, T.~J. and {Reichart}, D.~E. and
    {Berger}, E. and {Harrison}, F.~A. and {Price}, P.~A. and {Yost}, S.~A. and
    {Diercks}, A. and {Goodrich}, R.~W. and {Chaffee}, F.}, 2001. \apjl~
    562, L55

\bibitem[{Piran} {et~al.}(2001)]{Piran01}{Piran}, T. and {Kumar}, P. and {Panaitescu}, A. and
{Piro}, L., 2001, \apjl~ 560, L167.

\bibitem[{Panaitescu}, A. and {Kumar} (2001)]
{PK01}{Panaitescu}, A. and {Kumar}, P., 2001, \apjl~ 560, L49

\bibitem[{{Sedov}(1959)}]{Sedov59}
{Sedov}, L.~I. 1959, Similarity and Dimensional Methods in
Mechanics (New York:
  Academic Press)

\bibitem[{{Ayal \& Piran} (2001)}]{Ayal_Piran}
Ayal, S. \& Piran, T., 2001, \apj~ 555, 23


\bibitem[{{Woosley} (1993)}]{Woosley93}
S.~E. {Woosley}, \apj~ {\bf 405},  273  (1993)

\bibitem[{Paczynski} (1998)]{Pac98} B. {Paczynski}, \apjl~ {\bf 494},  L45  (1998).
\bibitem[{{MacFadyen and  Woosley} (1999)}]{MacFadyen_W99} A.~I. {MacFadyen} and S.~E. {Woosley}, \apj {\bf 524},  262
1999, \apj~ {\bf 524},  262

\bibitem[{{Ayal} {et~al.}(2001)}]{sphpn}
{{Ayal}, S. and {Piran}, T. and {Oechslin}, R. and {Davies},
M.~B. and    {Rosswog}, S.}, 2001, \apj~ 550, 846


\bibitem[{{Rhoads}(1997)}]{rhoads97}
{Rhoads}, J.~E. 1997, \apjl~ 487, L1


\bibitem[{{Lang}(1980)}]{lang80}
{Lang}, K.~R. 1980, Astrophysical Formulae (Springer-Verlag)

\bibitem[{{Shu}(1991)}]{shu91}
{Shu}, F.~H. 1991, The Physics of Astrophysics, Vol.~1
(University Science
  Books)


\bibitem[{{Williams} {et~al.}(1997){Williams}, {Chu}, {Dickel}, {Beyer},
  {Petre}, {Smith}, \& {Milne}}]{Williamsetal97}
{Williams}, R.~M., {Chu}, Y.~H., {Dickel}, J.~R., {Beyer}, R.,
{Petre}, R.,
  {Smith}, R.~C., \& {Milne}, D.~K. 1997, \apj~ 480, 618

\end{thebibliography}

\end{document}